\documentclass{jpsj2}

\def\runauthor{Online-Journal Subcommittee}

\title{%
Singular Vortex in Narrow Cylinders of Superfluid $^3$He-A Phase
}

\author{%
Yasumasa Tsutsumi\thanks{E-mail address: tsutsumi@mp.okayama-u.ac.jp} and Kazushige Machida
}

\inst{%
Department of Physics, Okayama University,
Okayama 700-8530
}

\recdate{\today}

\abst{%
Motivated by the on-going rotating cryostat experiments in ISSP, Univ. of Tokyo,
we explore the textures and vortices in superfluid $^3$He-A phase confined in narrow cylinders,
whose radii are $R$=50$\mu$m and 115 $\mu$m.
The calculations are based on the Ginzburg-Landau (GL) framework, which fully takes into
account the orbital ($\vec l$-vector) and spin ($\vec d$-vector) degrees of freedom for chiral $p$-wave pairing superfluid.
The GL free energy functional is solved numerically by using best known GL parameters 
appropriate for the actual experimental situations at $P$=3.2MPa and $H$=21.6mT.
We identify the ground state $\vec l$-vector configuration 
as radial disgyration (RD) texture with the polar core both at rest and low rotations
and associated $\vec d$-vector textures for both narrow cylinder systems under high magnetic fields.
The RD which has a singularity at center, changes into Mermin-Ho texture 
above the critical rotation speed which is determined precisely,
providing an experimental check for own proposal.
}

\kword{
$^3$He, A-phase, chiral $p$-wave pairing, superfluid, vortex,  texture
}

\begin{document}
\maketitle

\section{Introduction}

There has been much interest focused on superfluidity 
in various systems, ranging from superfluid $^3$He liquid, $^4$He super-solid\cite{chan}, neutral gases with bosonic 
($^{23}$Na, $^{87}$Rb, etc) and Fermionic atoms ($^{6}$Li)\cite{dalibard}, to superconductivity
with charged electrons, color superconductivity in dense quark matter\cite{casalbuoni}.
Among them, superfluid $^3$He\cite{dobbs,leggett,wolfle,volovik,salomaa,fetter-rev}
characterized by a triplet pairing occupied a special position because 
the rich internal degrees of freedom of a Cooper pair can be controlled by external nobs, such as 
the boundary condition of a confining wall, magnetic field and rotation.
These features of $^3$He have been demonstrated experimentally and theoretically,
providing us a testing ground whose theoretical basis is well established
to check various novel ideas and challenging proposals. 
Thorough studies in $^3$He as a prototype multi-component order parameter (OP) superfluid
are quite useful to understand other non-trivial pairing states, such as $p$-wave pairing superfluids via a Feshbach resonance, 
which is yet to be realized in ultracold atom gases\cite{tsutsumi}.
Possible application of this study may be to a heavy Fermion superconductor UPt$_3$, which consists of three 
superconducting phases in field $H$ and temperature $T$ plane\cite{machida}.
This phase diagram can not obviously be explained in terms of a single OP, but must be
multi-component OP. The high field and low $T$ phase, so-called C phase is analogous to
the present A-phase.

Superfluid $^3$He-ABM(A) phase stabilized at high pressures and high temperatures ($T$) over 
BW(B) phase can be described by the $\vec l$-vector for the orbital $p$-wave symmetry
and the  $\vec d$-vector for the spin symmetry. These two vectors fully characterize the spatial structure, i.e. texture
of the underlying OP symmetry\cite{dobbs,leggett,wolfle,volovik,salomaa,fetter-rev}.

It is known that in the absence of field, the Mermin-Ho (MH) texture\cite{MH,AT} is stable under confined geometries
at rest where spontaneous mass current flows along the boundary wall. This MH texture is so generic and 
stable topologically, thus characteristic  to multi-component OP superfluids. It exists even in  the spinor BEC\cite{mizushima,leanhardt,bigelow}.
However, as the confining system becomes smaller and comparable to their characteristic length scale $\sim \xi_d$ $(\simeq 10\mu$m),
the texture formation becomes difficult due to the kinetic energy penalty, leading to destabilization of MH.
Similarly MH becomes unstable under magnetic fields whose order is $H_d$ ($\simeq$2mT)\cite{fetter}. 
It is not known theoretically and experimentally what exactly condition needed for that and 
what kind of texture is stabilized in such a situation.

Here we are going to solve this problem in connection with
the ongoing experiments performed in ISSP, Univ. of Tokyo.
They use the two narrow cylinders with the half radii $R=50\mu$m and 115$\mu$m
filled with $^3$He-A phase.
These two kinds of cylinders are rotated up to the maximum rotation speed $\Omega$=11.5rad/sec
under  a field applied along the rotation axis $z$ and pressure $P$=3.2MPa.
They monitor the NMR spectrum to characterize textures created in
a sample. So far the following facts are found\cite{ishiguro,ishigurothesis,ishiguromeeting,izuminameeting,izumina}:

\noindent
(1) At rest the un-identified texture is seen for both samples with $R=50\mu$m and 115$\mu$m.
The texture, which has a characteristic resonance spectrum, persists down to the lowest temperature
T/T$_c$=0.7, below which the B phase starts to appear,
from the onset temperature T$_c=2.3$mK. Thus the ground state in those narrow cylinders is this 
un-identified texture.

\noindent
(2) Upon increasing $\Omega$ this texture is eventually changed into the MH texture which
was identified before for $R=115\mu$m by the same NMR experiment\cite{ishiguromeeting}.
This critical rotation speed $\Omega_c\sim 0.5$rad/sec, which is identified as a sudden intensity change 
of the main peak in the resonance spectrum.

\noindent
(3) With further increasing $\Omega$ the so-called continuous unlocked vortex (CUV) are identified
for $R=115\mu$m sample. The successive transitions from the MH to one CUV, two CUV, etc 
in high rotation regions are explained basically by Takagi\cite{takagi} who solves
the same GL functional as in the present paper. 
However, the calculations are assumed to be the A-phase in the whole region.
Thus the following calculation is consistent with those by Takagi\cite{takagi}.
Those successive transitions are  absent in $R=50\mu$m sample because the estimated critical
$\Omega$ for CUV exceeds the maximum rotation speed 11.5rad/sec in the rotating cryostat in ISSP.

\noindent
(4) The un-identified texture for $R=115\mu$m sample stable at rest and under low rotations
exhibits a hysteretic behavior about $\pm \bf \Omega$ rotation sense\cite{ishiguro}, meaning that 
this texture with ${\bf \Omega}>0$ differs from that with ${\bf \Omega}<0$, namely, this texture has polarity.
These textures with $\pm \bf \Omega$ can not continuously deform to each other.
This feature is completely absent for the texture for $R=50\mu$m sample
because the NMR spectra for $\pm \bf \Omega$ are identical\cite{izumina}.
Thus we have to distinguish two kinds of the texture stable at rest and under low rotations 
for $R$=50$\mu$m and 115$\mu$m.

\begin{figure}
\begin{center}
\includegraphics[width=7cm]{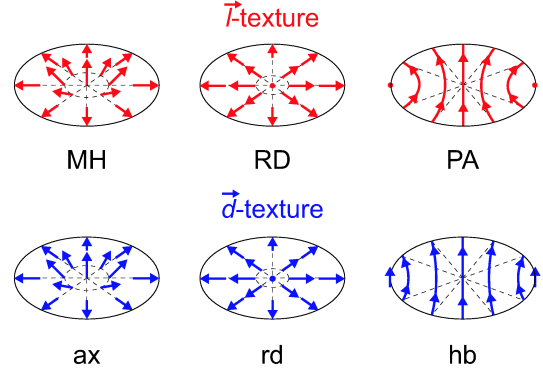}
\end{center}
\caption{(Color online) 
The possible textures of the  $\vec{l}$-vectors (upper) and $\vec{d}$-vectors (lower)
are shown schematically. The Mermine-Ho (MH), radial disgyration (RD), Pan-Am (PA)
are for the  $\vec{l}$-vector textures (upper panel). 
The axial type (ax), radial disgyration (rd)
and hyperbolic type (hb) are for the $\vec{d}$-vector textures (lower panel).
The points in MH, PA and rd-type denote positions of singularity.
}
\label{texture}
\end{figure}

Before going into the detailed calculations, we introduce here the possible textures
to be examined in the following: The texture consists of the orbital and spin parts,
each of which is characterized by the $\vec{l}$-vector and $\vec{d}$-vector respectively.
Thus the total order parameter, or the texture is fully characterized by the combination of the
$\vec{l}$-texture and $\vec{d}$-texture. 
In Fig. \ref{texture} we schematically show the possible $\vec{l}$-textures and $\vec{d}$-textures
where in MH the $\vec{l}$-vector and in ax-type the $\vec{d}$-vector change smoothly and flare out towards the wall.
In contrast, the RD\cite{deGennes} (rd-type) has a singularity at the center. 
The $\vec{l}$-vector in PA\cite{Maki} (the $\vec{d}$-vector in hb-type) shows
a hyperbolic like spatial structure. In the following we examine the four textures, MH-ax, MH-hb, RD-ax
and RD-hb, which turn out to be relevant and compete energetically each other in the present situations.
 
 The arrangement of the paper is as follows:
After giving the Ginzburg-Landau (GL) free energy functional, we set up 
the GL parameters appropriate for the present experimental situation in \textsection{} 2.
Here the boundary condition which is essential for the present narrow cylinders is examined.
We explain also our numerics to evaluate various possible textures.
In \textsection{} 3 we list up possible $\vec l$-vector and $\vec d$-vector textures both at rest and under rotation.
In \textsection{} 4 we identify the ground state texture by comparing the GL free energy
for various radii $R$ of the cylinder, rotation speeds  $\Omega$ and magnetic fields $H$.
We critically examine the on-going experiments at ISSP by using the rotating cryostat in the light
of the above calculations in \textsection{} 5. The final section is devoted to conclusion and summary.

\section{Formulation}
\subsection{Order parameter and GL functional}
The OP of superfluid $^3$He is given by a symmetric $2 \times 2$ matrix\cite{dobbs,leggett,wolfle,volovik,salomaa,fetter-rev}
\begin{align}
\hat{\Delta}(\hat{\bf p})=iA_{\mu i}\sigma_{\mu}\sigma_y\hat{p}_i,\quad (\mu,i=x,y,z),
\end{align}
where $\sigma$ is the Pauli matrix,
$\hat{\bf p}$ is the unit vector of the momentum on the Fermi surface.
The summations over repeated indices are implied.
Thus superfluid $^3$He is characterized rank-2 tensor components $A_{\mu i}$ 
inherent in the $p$-wave pairing ($L=1$) with spin $S=1$,
where $\mu$ and $i$ denote cartesian coordinates of the spin and orbital spaces, respectively.

In order to understand the stable texture at rest and lower rotations, we examine the standard GL
functional, which is well established thanks to the intensive theoretical and experimental studies over thirty years\cite{dobbs,leggett,wolfle,volovik,salomaa,fetter-rev}.
Namely, we start with the following GL form written 
in terms of the tensor $A$ forming OP of $p$-wave pairing. The most general GL functional density $f_{\rm bulk}$ 
for the bulk condensation energy up to fourth order is written as 
\begin{align}
f_{\rm bulk}=-\alpha A^*_{\mu i}A_{\mu i}+\beta_1A^*_{\mu i}A^*_{\mu i}A_{\nu j}A_{\nu j}
+\beta_2 A^*_{\mu i}A_{\mu i}A^*_{\nu j}A_{\nu j}
+&\beta_3 A^*_{\mu i}A^*_{\nu i}A_{\mu j}A_{\nu j}\nonumber \\
+\beta_4 A^*_{\mu i}A_{\nu i}A^*_{\nu j}A_{\mu j}
+&\beta_5 A^*_{\mu i}A_{\nu i}A_{\nu j}A^*_{\mu j},
\label{fbulk}
\end{align}
which is invariant under spin and real space rotations 
in addition to the gauge invariance U(1)$\times$SO$^{(S)}$(3)$\times$SO$^{(L)}$(3).
The coefficient $\alpha(T)$ of the second order invariant is $T$-dependent as usual,
and the fourth order terms have five invariants with coefficients $\beta_j$ in general.
The gradient energy consisting of the three independent terms is given by 
\begin{eqnarray}
f_{\rm grad}=K_1(\partial_i^* A_{\mu j}^* )(\partial_i A_{\mu j} )
+K_2(\partial_i^* A_{\mu j}^* )(\partial_j A_{\mu i} )
+K_3(\partial_i^* A_{\mu i}^* )(\partial_j A_{\mu j} ) .
\label{fgrad}
\end{eqnarray}
where $\partial_i=\nabla_i-i({2m_3\over \hbar})({\bf \Omega \times r})_i $ 
with the angular velocity $\bf \Omega\parallel z$, which is parallel to the cylinder long axis 
and ${\bf \Omega}>0$ means the counter clock-wise rotation.
In addition, there are the dipole and magnetic field energies:
\begin{align}
f_{\rm dipole}&=g_d\left( A_{\mu\mu}^*A_{\nu\nu} + A_{\mu\nu}^*A_{\nu\mu} 
- {2\over 3}A_{\mu\nu}^*A_{\mu\nu} \right),
\label{fdipole}\\
f_{\rm field}&=g_mH_{\mu}A_{\mu i}^*H_{\nu}A_{\nu i}.
\label{ffield}
\end{align}
In the A-phase the spin and orbital parts of the OP are factorized, 
i.e. $A_{\mu i}=\hat{d}_{\mu} A_i$,
where the spin part is denoted  by $\hat{d}_{\mu}$ and the orbital part by $A_i$.
The $\hat{\bf d}$ is a unit vector.
We can describe the ABM and polar state under this notation.
Now, Eqs.\eqref{fbulk}, \eqref{fgrad}, \eqref{fdipole} and \eqref{ffield} are rewritten as
\begin{align}
f_{\rm bulk}&=-\alpha A_i^* A_i +\beta_{13} A_i^* A_i^* A_j A_j +\beta_{245} A_i^* A_j^* A_i A_j, \\
f_{\rm grad}&=K_1(\partial_i^* \hat{d}_{\mu} A_j^* )(\partial_i \hat{d}_{\mu} A_j )
+K_2(\partial_i^* \hat{d}_{\mu} A_j^* )(\partial_j \hat{d}_{\mu} A_i )
+K_3(\partial_i^* \hat{d}_{\mu} A_i^* )(\partial_j \hat{d}_{\mu} A_j ) ,\\
f_{\rm dipole}&=g_d\left[ \hat{d}_{\mu}\hat{d}_{\nu} \left( A_{\mu}^*A_{\nu} +A_{\nu }^*A_{\mu } \right) 
-\frac{2}{3}A_{\nu }^*A_{\nu } \right],\\
f_{\rm field}&=g_m A_i^* A_i \left(\hat{\bf d} \cdot {\bf H} \right)^2,
\end{align}
where $\beta_{13}=\beta_1 +\beta_3$,  and $\beta_{245}=\beta_2 +\beta_4 +\beta_5$.

The coefficients $\alpha$, $\beta_j$, $K_j$, $g_m$, and $g_d$ 
are determined by Thuneberg\cite{Thuneberg1} and Kita\cite{Kita} as follows.
The weak-coupling theory gives
\begin{align}
\alpha = {N(0) \over 3}\left( 1-{T \over T_c} \right) \equiv \alpha_0 \left( 1-{T \over T_c} \right),
\label{GLa}
\end{align}
\begin{align}
-2\beta_1^{WC}=\beta_2^{WC}=\beta_3^{WC}=\beta_4^{WC}=-\beta_5^{WC}={7\zeta(3)N(0) \over 120(\pi k_B T_c)^2},
\end{align}
\begin{align}
K_1=K_2=K_3={7\zeta(3)N(0)(\hbar v_F)^2 \over 240 (\pi k_B T_c)^2} \equiv K,
\label{GLK}
\end{align}
where $N(0)$ and $v_F$ are the density of states per spin and the Fermi velocity, respectively.
The coefficients $\alpha_0$ and $K$ are estimated by Eqs. \eqref{GLa} and \eqref{GLK} 
by using the values of $N(0)$, $T_c$ and $v_F$ which are determined experimentally by Greywall\cite{Greywall} 
within the weak-coupling theory.
It is known that strong-coupling corrections for $\beta_j$ are needed to stabilize A-phase, and 
we use the $\beta_j$ values estimated by Sauls and Serene\cite{Sauls}.
The value of $g_d$ is\cite{Thuneberg2}
\begin{align}
g_d={\mu_0 \over 40} \left( \gamma \hbar N(0) \ln {1.1339 \times 0.45T_F \over T_c} \right)^2,
\end{align}
where $\mu_0$ and $\gamma$ denote the permeability of vacuum and the gyromagnetic ratio, respectively,
and $T_F$ is the Fermi temperature defined by $T_F \equiv 3n/4N(0)k_B$ with the density $n$.
Finally, $g_m$  is given within the weak-coupling expression by
\begin{align}
g_m={7\zeta(3)N(0)(\gamma \hbar)^2  \over 48\left[ (1+F_0^a)\pi k_B T_c \right]^2},
\end{align}
with $F_0^a$ the Landau parameter taken from Wheatley\cite{Wheatley}.

We set all the GL parameters to correspond to the above experimental pressure $P=3.2$MPa,
which are summarized as
\begin{align}
\alpha_0={N(0)\over 3}&=3.81\times 10^{50}{\rm J^{-1}m^{-3}}, \\
\beta_1&=-3.75\times 10^{99}{\rm J^{-3}m^{-3}}, \\
\beta_2&= 6.65\times 10^{99}{\rm J^{-3}m^{-3}}, \\
\beta_3&= 6.56\times 10^{99}{\rm J^{-3}m^{-3}}, \\
\beta_4&= 5.99\times 10^{99}{\rm J^{-3}m^{-3}}, \\
\beta_5&=-8.53\times 10^{99}{\rm J^{-3}m^{-3}}, \\
K&=4.19\times 10^{34}{\rm J^{-1}m^{-1}}, \\
g_d&=5.61\times 10^{44}{\rm J^{-1}m^{-3}}, \\
g_m&=1.35\times 10^{44}{\rm J^{-1}m^{-3}(mT)^{-2}}, \\
{2m_3 \over \hbar}&=9.51\times 10^{7}{\rm m^{-2}s}. 
\end{align}
Corresponding to  these parameters, the dipole field is estimated as $H_d=\sqrt{g_d/g_m}\sim 2$mT, 
which is a characteristic magnetic field where the dipole and magnetic field energies become same order.

The stable texture can be found by minimizing total free energy,
\begin{align}
F=\int d^3r f({\bf r}) = \int d^3r (f_{\rm bulk} + f_{\rm grad} + f_{\rm dipole} + f_{\rm field}).
\end{align}
We have identified stationary solutions by numerically solving the variational equations:
$\delta f({\bf r})/\delta \hat{d}_{\mu}({\bf r})=0$, $\delta f({\bf r})/\delta A_i({\bf r})=0$ 
in cylindrical systems, assuming the  uniformity towards the $z$ direction.
The height of the cylinders used in ISSP
experiments are $\sim$5mm, which allows us to adopt this assumption.
Thus we obtain the stable $\vec{d}$-textures and $\vec{l}$-textures. Namely,
we solve the coupled GL equations in two dimensions. 
The $\vec l$-vector is defined as
\begin{align} 
l_{i}\equiv-i\epsilon_{ijk}{A_j^*A_k\over |\Delta|^2},
\label{l-vector}
\end{align}
where $\epsilon_{ijk}$ is totally antisymmetric tensor
and $|\Delta|^2=A_i^*A_i$ is the amplitude of OP.
The mass current is given by
\begin{align} 
j_{i}\equiv{4m_3K \over \hbar}{\rm Im}(A_j^*\nabla_iA_j+A_j^*\nabla_jA_i+A_i^*\nabla_jA_j).
\end{align}
Here we expand the OP in a basis of spherical harmonics $Y_{lm}$ $(l=1, m=-1,0,1)$ 
for ease to express the initial configuration of the $\vec{l}$-textures;
\begin{align} 
\Delta ({\bf \hat{p}}) = A_+\hat{p}_+ + A_0\hat{p}_0 + A_-\hat{p}_-,
\end{align}
where $\hat{p}_{\pm} = \mp \left(\hat{p}_x \pm i\hat{p}_y \right)/\sqrt{2},\ \hat{p}_0=\hat{p}_z$ and
$A_{\pm} = \mp \left(A_x \mp iA_y \right)/\sqrt{2},\ A_0=A_z$.

It is noted that the characteristic length associated with the dipole energy
is given by $\xi_d=\sqrt{K/g_d}$, which is estimated to be an order of 10$\mu$m.
In contrast, the usual coherence length is $\xi=\sqrt{K/\alpha}=\xi_0 /\sqrt{1-T/T_c}$,
where the coherence length at zero temperature $\xi_0 \sim 0.01\mu$m.
These two length scales differ by three order of magnitude which causes great difficulty to handle the 
problem involved both scales simultaneously. This is indeed our problem.
The radial disgyration (RD) has a phase singularity at the center where the $\vec l$-vector
vanishes around $\xi$-scale region while MH has no singularity and $\vec l$-vector is non-vanishing everywhere
whose spatial variation is characterized by $\xi_d$. In order to compare two energies,
we need to handle two scales simultaneously. Since at the boundary $\vec l$-vector is 
constrained  such that it is always perpendicular to the wall, thus RD and MH exhibit a similar $\vec l$-vector texture, 
differing only around the center of a cylinder.
We evaluate possible $\vec l$-textures RD and MH in combination with $\vec d$-textures; axial type (ax)
and hyperbolic type (hb). 
The radial disgyration of $\vec d$-texture (rd) is neglected since it has a singularity where superfluidity is broken.
Namely, we mainly examine four kinds of texture, RD-ax, RD-hb, MH-ax and MH-hb 
in addition to the Pan-Am (PA) only for smaller systems in this paper. Those textures, we believe,  exhausts all relevant
stable textures. There is no other texture known in literature\cite{wolfle}.

\subsection{Numerics and boundary condition}

The numerical computations have been done by using
polar coordinates $(r, \theta)$. The radial direction $r$ is discretized into 1000 meshes for R=50$\mu$m system
and 2300 meshes for R=115$\mu$m while the azimuthal angle $\theta$ is discretized in 180 points.
Thus the total lattice points are 1000$\times$180 and 2300$\times$180 in $(r, \theta)$ coordinate system
for R=50$\mu$m and R=115$\mu$m respectively.
The average lattice spacing is an order of 5$\xi_0$ for both cases, which is fine enough to accurately 
describe a singular core in RD. Note that we are considering high temperature region, 
$\xi(T=0.95T_c)\sim 4.5\xi_0\sim 45$nm
while our lattice spacing is 50nm. The advantages of using $(r, \theta)$ coordinate system over
the rectangular $(x,y)$ system are (1) we can reduce the total lattice points, keeping the numerical accuracy
and (2) it is easy to take into account the boundary condition at the wall where the $\vec l$-vector 
orients along the perpendicular direction to the wall, namely the radial direction $\bf r$.
The disadvantage is that we can not describe the Pan-Am (PA) type configuration for the $\vec l$-vector.
However, PA is topologically similar to RD because as seen from Fig. \ref{texture} two singularities at the wall in 
PA merge into a singularity  at the center in RD. Thus we may ignore PA and concentrate on RD only
except for smaller radius cases $(R\leq 20\mu$m) where we confirm the above statement (see subsection 3.3 for detail).

On each lattice points the 9 variational parameters are assigned, coming from the complex variable
$A_i (i=x,y,z)$ and the real three dimensional vector $d_{\mu}(\mu=x,y,z)$.
These 9 parameters are determined iteratively and self-consistently by solving 
the coupled GL equations. One solution is needed $\sim$7 days for R=50$\mu$m system
and $\sim$20 days for R=115$\mu$m system by using OpenMP programming on a XEON 8 core machine.
The applied field $H=21.6$mT is taken for all calculation except for subsection 4.2.

\section{Stable Textures}

In order to help identifying the possible texture realized in narrow cylinders ($R$=50$\mu$m and 115$\mu$m)
both at rest and under rotation, we first examine the detailed spatial structures for each texture, which consists of
the $\vec l$-vector and $\vec d$-vector before discussing the relative stability among them
under actual experimental setups.
Here we study mainly four types of the textures: RD-hb, RD-ax, MH-hb and MH-ax, which are relevant
for later discussions on experiments.
As a supplement, we also examine PA-hb for smaller sizes.

\subsection{Radial disgyration (RD)}

The RD is characterized by having a singularity at the center where $A_{\pm}(r=0)=0$ and $A_0(r=0)\neq 0$.
Thus the vortex core is filled by $A_0$ component, namely it is a polar core vortex\cite{ikeda}. 
The associated $\vec d$-vector texture could be either hb-type or ax-type. 
The $\vec l$-vector texture can be obtained by starting with an initial configuration:
\begin{align}
A_+(r, \theta)&={\Delta_A\over 2}{\tanh}({r\over\xi})e^{-i\theta}, \nonumber\\
A_0(r, \theta)&={\Delta_A \over \sqrt{2}}, \\
A_-(r, \theta)&={\Delta_A\over 2}{\tanh}({r\over\xi})e^{i\theta}\nonumber,
\end{align}
where $\Delta_A$ is the amplitude of OP in the bulk region far away from the center and boundary.
The combination of the winding number in RD is $(w_+,w_0,w_-)=(-1,0,1)$ where $w_+,w_0,w_-$
are the winding numbers for $A_+,A_0,A_-$ respectively.
The $A_{\pm}$ components vary over $\xi$ distance from the center
while the $A_0$ component stays constant throughout the system.
Since at the center only the $A_0$ component is non-vanishing, the polar state is realized there as mentioned before.
It is seen from Eq. \eqref{l-vector} that in this RD form $l_z=0$ because of $|A_+|=|A_-|$.

\begin{figure}
\begin{center}
\includegraphics[width=14cm]{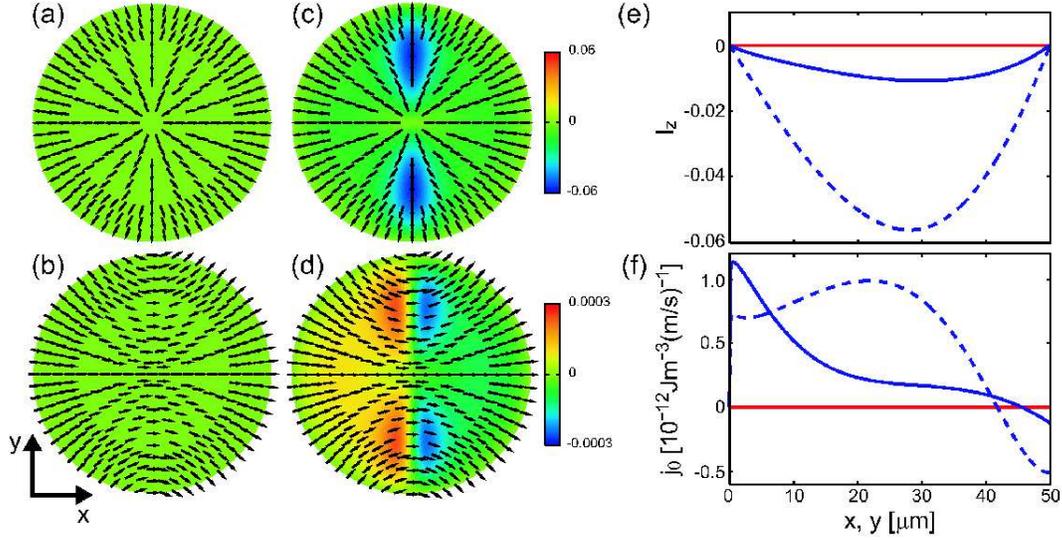}
\end{center}
\caption{(Color online) 
Stable textures and current for RD-hb under $R=50\mu$m and $T/T_c=0.95$.
(a) $\vec{l}$-vector textures at rest.
(b) $\vec{d}$-vector textures at rest.
(c) $\vec{l}$-vector textures at $\Omega=3$rad/s.
(d) $\vec{d}$-vector textures at $\Omega=3$rad/s.
The arrows indicate the $x$-$y$ component and color codes denote
its $z$ component.
(e) The $z$ component of the $\vec{l}$-vector and 
(f) the current $j_{\theta}$ along the radial direction $r$ from the center.
The red lines are at rest and blue lines are $\Omega=3$rad/s.
The solid (dashed) lines are along $x$ ($y$) axis.
}
\label{RD-hb}
\end{figure}

In Fig. \ref{RD-hb} we show the results of the stable RD-hb textures for R=50$\mu$m,
where Figs. \ref{RD-hb}(a) and \ref{RD-hb}(b) at rest and Figs. \ref{RD-hb}(c) and \ref{RD-hb}(d) at $\Omega=3$rad/s.
It is seen that $\vec l$-vectors flare out from the center and point perpendicular to the wall (Fig. \ref{RD-hb}(a)).
The $\vec d$-vectors point almost to the horizontal direction (the $x$-axis), 
and curve near the wall (Fig. \ref{RD-hb}(b)),
because the dipole interaction tends to align  the $\vec d$-vector parallel to the $\vec l$-vector direction.
We call it ``hyperbolic'' (hb). 
The $\vec l$-vectors do not point completely to the radial direction,
but are twisted slightly due to the dipole interaction.
Since the $\vec l$-vectors are strongly constrained by the boundary condition,
they should be perpendicular to the wall
so as to suppress the perpendicular motion of the Cooper pair.
In other words, one of the two point nodes which locate to the $\vec l$-vector direction
should direct to the wall, thus saving the condensation energy loss.
Note that there is no $z$-components for both $\vec l$ and $\vec d$ in RD-hb at rest.
It will be soon shown that this RD  texture is most stable at rest and lower rotations
in narrow cylinders.

Once the rotation is turned on, the $l$-vectors and $\vec{d}$-vectors now acquire the $z$ components.
It is seen from Fig. \ref{RD-hb}(c) that  the $\vec{l}$-vectors tend to point the negative $z$ direction, 
recognized as the color changes and also seen from Fig. \ref{RD-hb}(e).
That region is confined along the $y$-axis where the dipole interaction, which tend to $\vec{l} \parallel \vec{d}$, 
is not effective compared with other regions.
In Fig. \ref{RD-hb}(e) the cross sections of $l_z$ component are displayed.
It is clear that $l_z$ component is mainly induced along the $y$ axis.
Simultaneously, 
as shown in Fig. \ref{RD-hb}(d) the $\vec{d}$-vectors point to $\pm z$ directions whose boundary occurs along 
the vertical direction, that is, the $y$-axis.
As shown in Fig. \ref{RD-hb}(f) the variation of the $\vec{l}$-vectors produces the in-plane mass current.
The $j_{\theta}(x,y)$  distributions in the $x$ and $y$ plane are anisotropic.
Along the radial direction the $j_{\theta}$ changes the sign, namely,
it is the same direction as the external rotation in the central region while opposite in the outer region.

\begin{figure}
\begin{center}
\includegraphics[width=14cm]{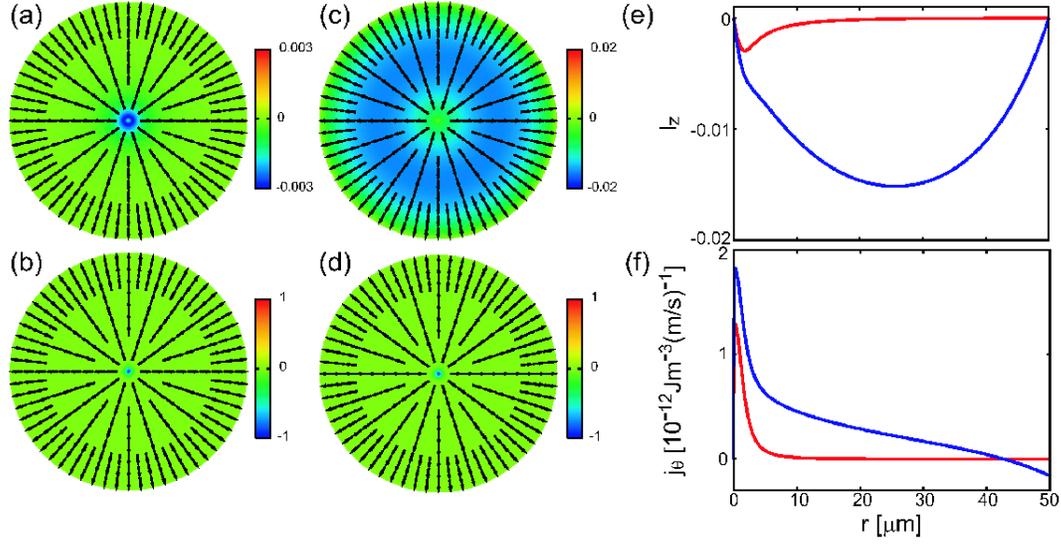}
\end{center}
\caption{(Color online) 
Textures and current for RD-ax under $R=50\mu$m and $T/T_c=0.95$.
(a) $\vec{l}$-vector textures at rest.
(b) $\vec{d}$-vector textures at rest.
(c) $\vec{l}$-vector textures at $\Omega=3$rad/s.
(d) $\vec{d}$-vector textures at $\Omega=3$rad/s.
The arrows indicate the $x$-$y$ component and color codes denote
its $z$ component.
(e) The $z$ component of the $\vec{l}$-vector and (f)
the current $j_{\theta}$ along the radial direction $r$ from the center.
The red lines are at rest and blue lines are at $\Omega=3$rad/s.
}
\label{RD-ax}
\end{figure}

As for RD-ax, which is displayed in Fig. \ref{RD-ax}. 
The $\vec l$-vector and $\vec d$-vector textures at rest are shown in Figs. \ref{RD-ax}(a) and \ref{RD-ax}(b), 
respectively.
These textures are seen to be cylindrically symmetric.
The $\vec d$-vectors around the center point to the negative $z$ direction (Fig. \ref{RD-ax}(b)).
The area is characterized by magnetic coherence length $\xi_h \equiv \sqrt{K/g_mH^2}$,
which is estimated to be an order of 1 $\mu$m in $H=21.6$mT.
This length scale is larger than $\xi$, but smaller than $\xi_d$.
Through the dipole interaction the $z$ component of the $\vec l$-vector is induced in that area 
(see Fig. \ref{RD-ax}(a) and red line in Fig. \ref{RD-ax}(e)).
The mass current flows even at rest exclusively around the center (see the red line in Fig. \ref{RD-ax}(f)).
Under rotation the $z$ component of the $\vec l$-vector in  RD-ax acquires more negative component in the whole region
(Fig. \ref{RD-ax}(c) and blue line in Fig. \ref{RD-ax}(e)),
so that the mass current increases further (blue line in Fig. \ref{RD-ax}(f)).
The $\vec d$-texture almost remains unchanged (Fig. \ref{RD-ax}(d)). 

\subsection{Mermin-Ho (MH)}

The MH texture does not include singularities, being the ABM state throughout the system.
The $\vec l$-vector is directed to the $z$ axis at the center, avoiding singularities.
An initial configuration for the $\vec l$-vector texture is
\begin{align}
A_+(r,\theta)&={\Delta_A \over 2}[1+\cos \beta(r)], \nonumber\\
A_0(r,\theta)&={\Delta_A \over \sqrt{2}}\sin \beta(r)e^{i\theta}, \\
A_-(r,\theta)&={\Delta_A \over 2}[1-\cos \beta(r)]e^{2i\theta}, \nonumber
\end{align}
where $\beta(r)=\pi r/2R$ varies linearly from 0 at the center to $\pi /2$ at the wall.
Thus the $\vec l$-vector points towards the $z$ direction at the center and the radial direction at the wall.
The $\vec d$-vector texture could be either hb-type or ax-type.

\begin{figure}
\begin{center}
\includegraphics[width=10cm]{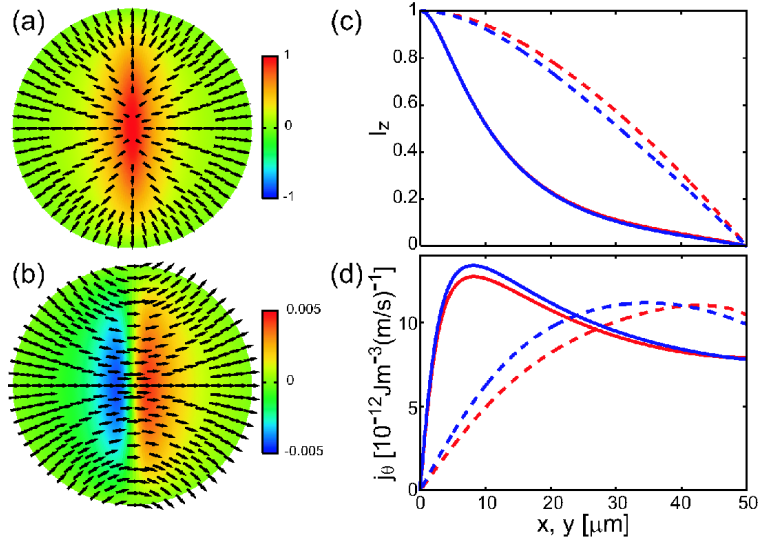}
\end{center}
\caption{(Color online) 
Stable textures and current for MH-hb under $R=50\mu$m and $T/T_c=0.95$.
(a) $\vec{l}$-vector and (b) $\vec{d}$-vector textures at rest.
The arrows indicate the $x$-$y$ component and color codes denote
its $z$ component.
(c) The $z$ component of the $\vec{l}$-vector and (d) 
the current $j_{\theta}$ along the radial direction $r$ from the center.
The red lines are at rest and blue lines are $\Omega=3$rad/s.
The solid (dashed) lines are along $x$ ($y$) axis.
}
\label{MH-hb}
\end{figure}

\begin{figure}
\begin{center}
\includegraphics[width=7cm]{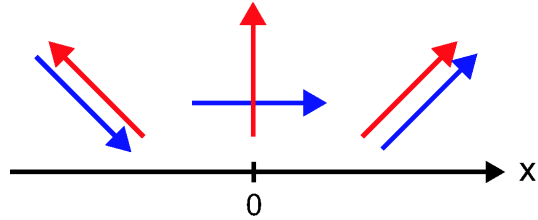}
\end{center}
\caption{(Color online) Arrangements of the $\vec l$-vectors (red arrows) and $\vec d$-vectors (blue arrows) 
for the MH-hb are shown schematically.
The three pairs of arrows show schematically the variations along the $x$ axis around the center.
Note that the angle between the $\vec d$-vectors and the $x$ axis is exaggerated.
}
\label{l-d}
\end{figure}

The MH-hb is shown in Fig. \ref{MH-hb} both at rest and under rotation.
At rest the $z$ component of the $\vec l$-vectors is non-vanishing around the center in an anisotropic manner 
(as seen from Fig. \ref{MH-hb}(a)).
The anisotropy of the $l_z$ component in the $x$-$y$ plane is understood in terms of the dipole interaction,
which is shown in Fig. \ref{MH-hb}(c).
The correlation between the $\vec l$-texture and $\vec d$-texture 
produces the anisotropic arrangements in each textural configuration.
The elongated $l_z$ component along the $y$ direction is already explained in the previous subsection.
The length scales of the $l_z$ variation towards the $x$ and $y$ directions 
are characterized by $\xi_d$ and $R$ respectively.
The $\vec l$-vector and the associated $\vec d$-vector variations along the $x$ axis at $y=0$ around the center 
are shown schematically in Fig. \ref{l-d}.
It is seen that since the $\vec l$-vectors (red arrows) flare out from the center,
the $z$ component of the $\vec d$-vector  (blue arrows) changes the sign at $x=0$.
This configuration embedded in the hyperbolic $\vec d$-texture is most advantageous by increasing the parallel portion 
$(\vec{l} \parallel \vec{d})$ due to the dipole interaction.
Since this structure is characterized by the winding number combination $(w_+,w_0,w_-)=(0,1,2)$,
it yields the spontaneous mass current at rest (red lines in Fig. \ref{MH-hb}(d)).
Thus MH has polarity which breaks the symmetry for $\pm \bf \Omega$ under rotation,
and  $j_{\theta}(x,y=0)$ is larger than $j_{\theta}(x=0,y)$ around the center,
and this relation is reversed in the outer region.
Under rotation the overall configuration of the $\vec l$-vector and $\vec d$-vector textures are not much changed 
compared with those at rest.
The small changes of the $l_z$ component and $j_{\theta}$ are shown in Figs. \ref{MH-hb}(c) and \ref{MH-hb}(d).

\begin{figure}
\begin{center}
\includegraphics[width=10cm]{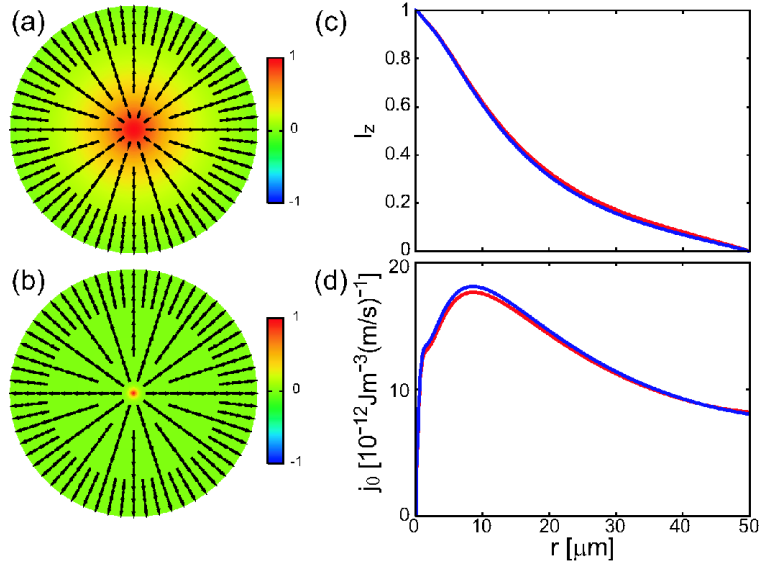}
\end{center}
\caption{(Color online) 
Textures and current for MH-ax under $R=50\mu$m and $T/T_c=0.95$.
 (a) $\vec{l}$-vector and (b) $\vec{d}$-vector textures at rest.
The arrows indicate the $x$-$y$ component and color codes denote
its $z$ component.
(c) The $z$ component of the $\vec{l}$-vector and (d) 
the current $j_{\theta}$ along the radial direction $r$ from the center.
The red lines are at rest and blue lines are $\Omega=3$rad/s.
}
\label{MH-ax}
\end{figure}

As for MH-ax shown in Fig. \ref{MH-ax},
the overall textures for $\vec l$-vector and $\vec d$-vector have cylindrical symmetry.
Note that the winding number combination $(0,1,2)$ is the same as before.
The $\vec l$-vectors and $\vec d$-vectors around the center 
whose length scales are characterized by $\xi_d$ and $\xi_h$, respectively.
These two length scales also appear in $j_{\theta}$ as clearly seen from Fig. \ref{MH-ax}(d),
where the sharp rise corresponds to $\xi_h$ and the maximum position to $\xi_d$.
The $l_z$ component in Fig. \ref{MH-ax}(c) shows gradual changes characterized by $\xi_d$.
The external rotation hardly change these features.

\subsection{Pan-Am texture}

\begin{figure}
\begin{center}
\includegraphics[width=7cm]{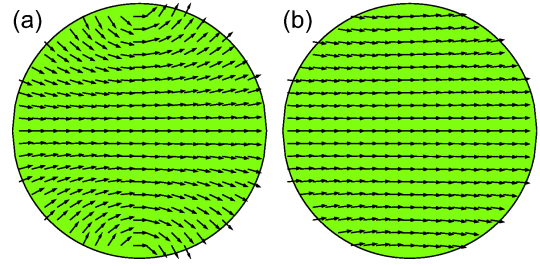}
\end{center}
\caption{(Color online) 
(a) $\vec{l}$-vector and (b) $\vec{d}$-vector textures
at rest for PA-hb under $R=20\mu$m and $T/T_c=0.95$.
The arrows indicate the $x$-$y$ component.
}
\label{PA-hb}
\end{figure}

The $\vec l$-texture in Pan-Am (PA) is characterized by the two singularities at the wall as
shown schematically in Fig. \ref{texture}. 
Here we show the stable PA texture for smaller size system R=20$\mu$m in Fig. \ref{PA-hb}.
In order to stabilize the PA we start with the initial configuration given by
\begin{align}
A_+(r, \theta)&=-{\Delta_A\over 2}(\sin\alpha-i\cos\alpha), \nonumber\\
A_0(r, \theta)&=i{\Delta_A \over \sqrt{2}}, \\
A_-(r, \theta)&={\Delta_A\over 2}(\sin\alpha+i\cos\alpha), \nonumber
\end{align}
with
$$\tan\alpha(r, \theta)=-{r^2\sin2\theta\over {R^2+r^2\cos2\theta}},$$
where $R$ is the radius of the system. 
The angle $\tan \alpha$ diverges at $(r,\theta)=(R,\pm \pi/2)$.
At those positions the OP amplitudes $A_{\pm}$ must be zero so that those are avoided.
Around those points the phase windings are $(w_+,w_0,w_-)=(-1,0,1)$.
Therefore those are analogous to the RD in the phase structure because
the singularity at the center in RD is regarded to be split into two in PA.

We show here the stabilized PA structure in Fig. \ref{PA-hb} where the $\vec l$-vector
and $\vec d$-vector structures are displayed.
From Fig. \ref{PA-hb}(a) the $\vec l$-vectors tend to curve near the wall to satisfy the boundary condition,
which causes the curving of the $\vec d$-vectors shown in Fig. \ref{PA-hb}(b),
otherwise they are straight pointing to the $x$ direction.
We notice that the energy of this PA is higher than the previous four kinds of the textures mentioned above.
This is also true for other systems with different $R$'s.
In the following discussions we disregard the PA,
only considering the previous four textures.

\section{Size-Dependence and Magnetic Field Effects}

\subsection{Size-dependence and the critical rotation speed}

\begin{figure}
\begin{center}
\includegraphics[width=7cm]{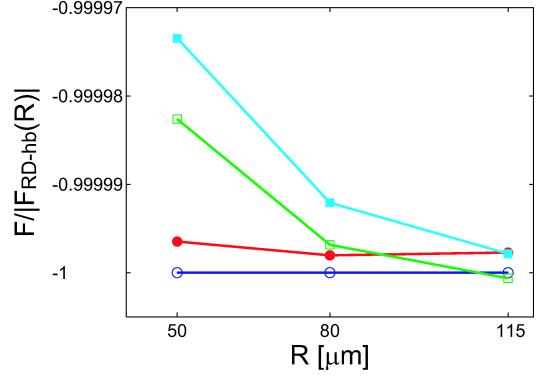}
\end{center}
\caption{(Color online) 
Size dependences of the free energies for various textures at rest.
For $R$=50$\mu$m RD-hb (blue) is stable while for $R$=115$\mu$m RD-ax (green)
becomes stable. 
The other lines show MH-hb (red) and MH-ax (light blue).
}
\label{R-F}
\end{figure}

We show the free energy of four textures, which are RD-hb (blue), RD-ax (green), MH-hb (red) and MH-ax (light blue), 
in Fig. \ref{R-F} for $R=50\mu$m, $R=115\mu$m and intermediate radius systems at rest.

The most stable texture for the $R=50\mu$m system is the RD-hb.
The reasons are understood by the following.
First we compare the free energy of the RD-hb with the MH-hb.
The RD-hb is advantageous from the view point of the dipole energy 
because the $\vec l$-vectors lie in the $x$-$y$ plane together with the $\vec d$-vectors.
The winding numbers of the RD-hb are $(w_+,w_0,w_-)=(-1,0,1)$,
so that the superflow velocity due to the phase gradient of the plus and the minus components is canceled.
In contrast, since the winding numbers of the MH-hb are $(w_+,w_0,w_-)=(0,1,2)$, 
the MH-hb has the spontaneous current comprising of the superflow velocity.
Consequently the MH-hb costs the gradient energy at rest.
On the other hand, the RD-hb costs the condensation energy
because the polar core with the radius $\xi$ exists at the center of the cylinder.
Nevertheless, since the polar core is small enough compared with the system size,
the RD-hb is more stable than the MH-hb.

Next we compare the free energy of the RD-hb with the RD-ax.
The gradient energy is favorable for the RD-hb since the $\vec d$-vectors are almost uniform except near the wall.
In addition, the magnetic field energy is gained
because the $\vec d$-vectors lie in the $x$-$y$ plane on application of the magnetic field toward $z$ direction.
On the other hand, the RD-ax is gained by the dipole energy
because the $\vec d$-vectors are parallel to the $\vec l$-vectors 
except with the radius $\xi_h$ around the center of the cylinder.
Nevertheless, 
since the dipole-unlocked (the $\vec l$-vectors do not parallel the $\vec d$-vectors) regions are small for the RD-hb, 
as well, the RD-hb is more stable than the RD-ax.
For the same reason, the MH-hb is more stable than the MH-ax,
and also the RD-hb is more stable than the MH-ax.
Therefore the RD-hb is the most stable texture for the $R=50\mu$m system at rest.

The most stable texture becomes the RD-ax for the $R=115\mu$m system.
The reason is that the dipole-unlocked regions enlarge by increasing the system size for the RD-hb,
whereas the regions of the RD-ax characterized by the magnetic coherence length $\xi_h$ are almost constant.
That is, the RD-ax is stabilized by the dipole energy for larger systems.
Similarly the stability between the MH-ax and the MH-hb is interchanged.

Under rotation, the MH textures are stable because they have the spontaneous current 
and they are tolerable under rotation compared with the RD textures which have very little spontaneous current.
The external rotation drives the RD-hb into the MH-hb for the $R$=50$\mu$m system.
Similarly the most stable texture RD-ax for the $R=115\mu$m system at rest changes into MH-hb by the rotation.
Therefore $\vec l$-texture and $\vec d$-texture change at the critical rotation speed.

\begin{figure}
\begin{center}
\includegraphics[width=7cm]{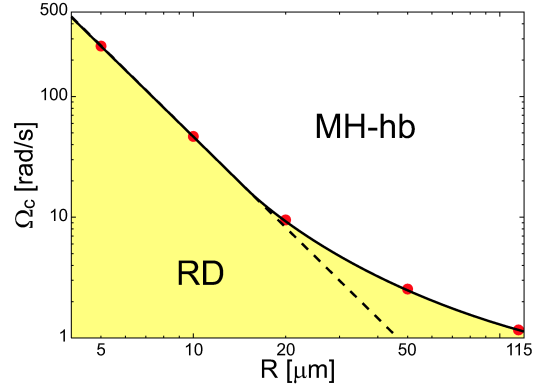}
\end{center}
\caption{(Color online) 
Critical rotation speed $\Omega_c$[rad/s] versus $R$[$\mu$m]
between RD and MH.
The dashed line is proportional to $R^{-2.5}$.
RD is most stable than MH at rest and low rotations for various
system sizes, including $R$=50$\mu$m and 115$\mu$m.
}
\label{R-Omega}
\end{figure}

The rotational speed changing the most stable $\vec l$-texture from the RD to the MH 
is defined as critical rotation speed $\Omega_c$.
The $\Omega_c$ obtained from the numerical calculations for the systems with the various radii $R$ 
is shown as red solid circles in Fig. \ref{R-Omega}.
The RD is the most stable $\vec l$-texture from the $R=5\mu$m to the $R=115\mu$m systems at rest,
and the MH-hb is the most stable texture under rotation.
The $\Omega_c$ decreases with increasing the radius of the systems.
The numerical results of the $\Omega_c$ curve upward than the dashed line, which is proportional to $R^{-2.5}$.
Therefore it is concluded that the RD is the most stable texture up to the $R\rightarrow \infty$ systems at rest.
Note that the results are obtained under the high magnetic field.
It has been pointed out by Buchholtz and Fetter\cite{fetter} 
that as the radius $R$ of the systems is larger, the RD is more favorable under high magnetic fields.

\subsection{Magnetic field effect}

\begin{figure}
\begin{center}
\includegraphics[width=7cm]{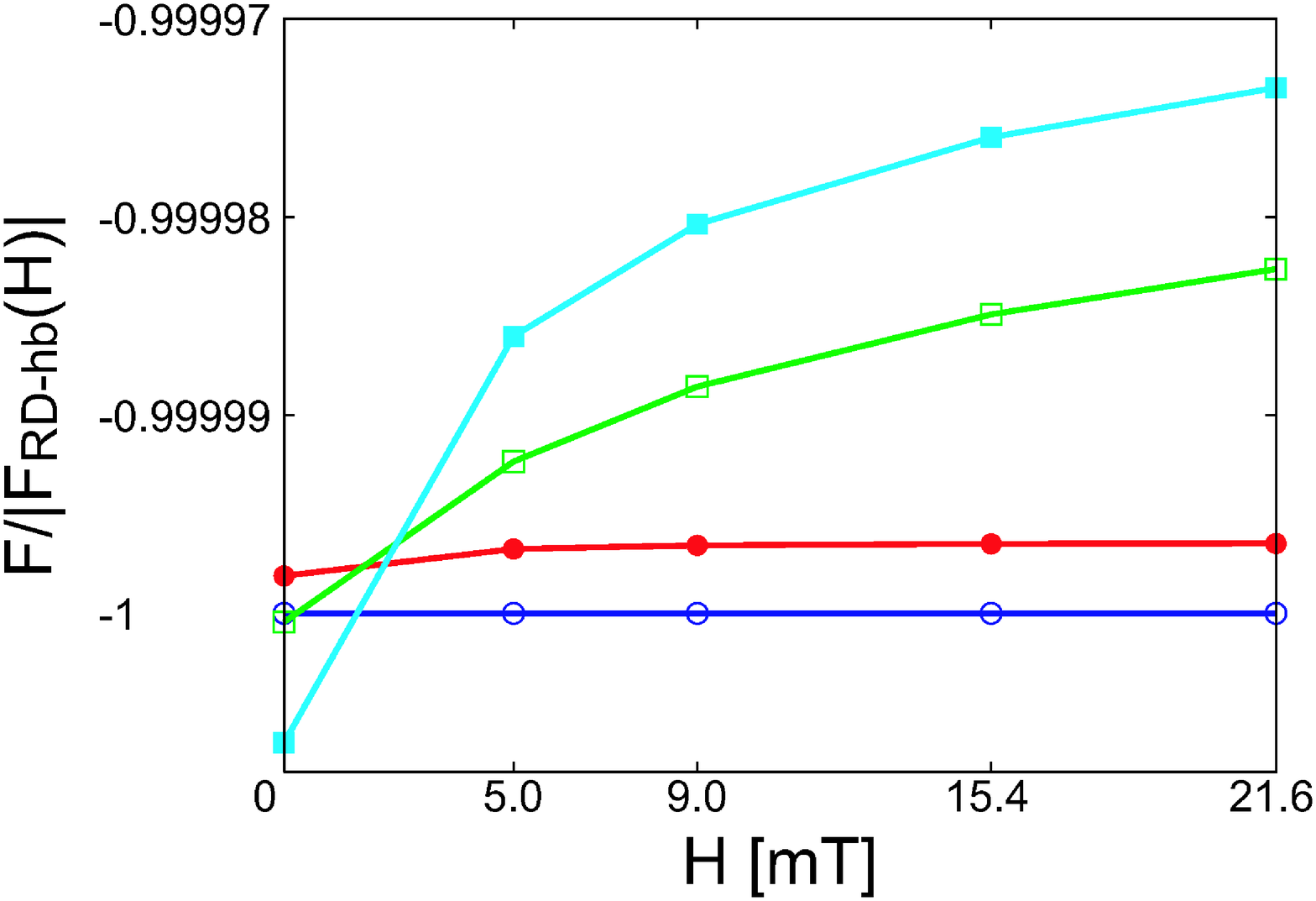}
\end{center}
\caption{(Color online) 
Comparison of various texture free energies as a function of the magnetic field $H$.
The switching between RD-hb (blue) and MH-ax (light blue) occurs at $H$=2mT.
The other lines show RD-ax (green) and MH-hb (red).
}
\label{H-F}
\end{figure}

So far we have fixed the magnetic field at $H$=21.6mT, which is selected by experiments
at ISSP because the MNR signals are best resolved.
From the theoretical point of view it is interesting to know in what weaker field MH is realized over RD.
We show the result in Fig. \ref{H-F} where various textures are compared as a function of $H$
for $R$=50$\mu$m and $T/T_c$=0.95.
It is found that at $H$=2mT MH-ax becomes stable over RD-hb beyond which RD-hb is always stable.
Below the dipole magnetic field MH is
expected to be stable over RD although it might be difficult to obtain good NMR signals in such a weak field.
Thus for feasible magnetic field region around $H$=21.6mT where sensible NMR signal is detectable, RD-hb is always stable
in the narrow cylinders.

\section{Analysis of Experiments}

\subsection{R=50$\mu$m}

\begin{figure}
\begin{center}
\includegraphics[width=7cm]{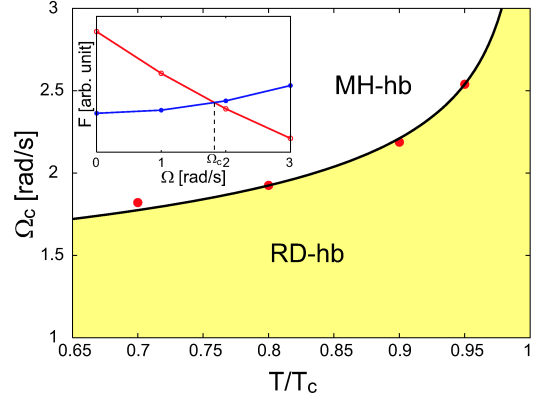}
\end{center}
\caption{(Color online) 
Phase diagram of critical rotation speed $\Omega_c$ (boundary line) and $T/T_c$ ($R=50\mu$m).
At rest RD-hb is always stable. The inset shows an example of the energy crossing between 
RD-hb (blue) and MH-hb (red) at $T/T_c=0.7$, showing $\Omega_c=1.8$rad/s.
}
\label{T-Omega}
\end{figure}

In Fig. \ref{T-Omega} we plot the critical rotation speed $\Omega_c$ as a function of temperature $T/T_c$
for $R$=50$\mu$m case. At rest RD-hb is always stable over MH-hb.
Upon increasing $\Omega$,  MH-hb becomes stable at $\Omega_c$, which
is an increasing function of $T/T_c$ because the vortex core energy $\sim \ln R/\xi$
becomes lower as approaching $T_c$ where $\xi$ is longer.
The reason why MH-hb is advantageous over RD-hb under rotation is that 
MH-hb has the spontaneous current, in contrast, RD-hb has no spontaneous current.
These characteristics are seen in the inset of Fig. \ref{T-Omega}.

NMR spectroscopy can be used to identify different topological objects of rotating superfluid $^3$He-A phase\cite{Ruutu}, 
namely it can be used to identify texture in a narrow cylinder.
The dipole-locked ($\vec{l} \parallel \vec{d}$) and $\vec{d} \perp {\bf H}$ regions  
occupying most of the condensates in the cylinder yields the main peak of NMR spectra.
In the high field limit its frequency is given by\cite{dobbs,leggett,wolfle}
\begin{align}
\omega_t^2 = \omega_L^2 + \omega_l^2.
\end{align} 
where $\omega_L =\gamma H$ is the Larmor frequency 
and $\omega_l$ is the longitudinal resonance frequency of the $^3$He-A phase.
The regions where the $\vec d$-vectors are dipole-unlocked and deviate from the
perpendicular direction to the field, generate the small satellite peaks.
The resonance frequency of a satellite is expressed in terms of a relative frequency shift $R_t^2$, defined by the equation
\begin{align}
\omega_t^2 = \omega_L^2 + R_t^2\omega_l^2.
\end{align} 
Therefore the texture in a narrow cylinder is identified by finding $R_t^2$.

In experiment of the NMR spectroscopy for various temperatures 
a satellite peak with $R_t^2 \simeq 0.8$ is observed at rest\cite{izuminameeting}.
The value of $R_t^2$ corresponds approximately to that by the RD-hb\cite{takagi2}.
We also notice that since RD-ax, 
which has a small area generating satellite peaks around the center of the cylinder with a radius $\xi_h$,
gives $R_t^2 \simeq 1.0$, the realized texture is definitely RD-hb, not RD-ax.
Upon increasing $\Omega$, RD-hb changes into MH-hb.
This texture has been identified previously by Takagi\cite{takagi} who has calculated $R_t^2 \simeq 0.3$ for MH-hb.

The precise determination of the temperature dependence of $\Omega_c(T)$ is under way experimentally.
Thus it is unable to check our prediction shown in Fig. \ref{T-Omega} at this time.
We emphasize here, however, that the NMR spectra at low $\Omega$ and high $\Omega$ are distinctively different,
therefore it is clear that the phase transition between two textures RD-hb and MH-hb occurs.

Experimentally\cite{izumina} there is no hysteresis for $\pm {\bf \Omega}$,
indicating that the texture at rest has no polarity to the direction of rotation.
This is in agreement with our identification of RD-hb for $R$=50$\mu$m, which has no polarity.

\subsection{R=115$\mu$m}

\begin{figure}
\begin{center}
\includegraphics[width=7cm]{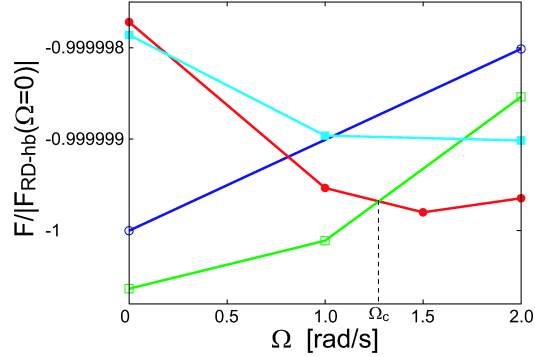}
\end{center}
\caption{(Color online) 
Free energy comparison of various textures for $R=115\mu$m.
RD-ax (green) changes to MH-hb (red) at $\Omega_c=1.3$rad/s. 
The other lines show RD-hb (blue) and MH-ax (light blue).
}
\label{Omega-F}
\end{figure}

As shown in previous section,
for the system size $R$=115$\mu$m the ground state texture at rest is RD-ax which is stabler than MH-hb.
The critical rotation speed $\Omega_c$=1.3rad/s at $T/T_c$=0.95 is calculated as shown in Fig. \ref{Omega-F}.
The value of $\Omega_c$ roughly coincides with the ISSP experiment\cite{ishiguromeeting}.

Experimentally there is a large hysteresis behavior about $\pm {\bf \Omega}$ centered at $\Omega$=0\cite{ishiguro},
which is contrasted with $R=50\mu$m case mentioned above.
This is understood because RD-ax has polarity where
$\vec l$-vectors and $\vec d$-vectors point to one of the directions $\pm z$-axis around the center which breaks
$\pm {\bf \Omega}$ symmetry.

As for the NMR spectrum at rest, they observe no distinctive satellite feature for $R$=115$\mu$m\cite{izuminameeting}.
These facts do not contradict our identification of RD-ax, which is $R_t^2\sim 1.0$.
Thus we expect no satellite feature for RD-ax.

\section{Conclusion and Summary}
\subsection{Conclusion}
Let us discuss the experimental facts (1)-(4) introduced in \textsection{} 1.

\noindent
(1) We have identified the unknown texture as RD-hb for 50$\mu$m sample and RD-ax for
115$\mu$m sample at rest and lower rotations by finding the most stable texture in those 
conditions.

\noindent
(2) Upon increasing the rotation, we found that RD changes into MH
where MH was identified before by NMR resonance shape unambiguously\cite{ishiguro}
for 115$\mu$m sample under rotations. This is confirmed by the present calculation.
Namely, we precisely determined the critical rotation speed $\Omega_c$ from RD to MH
for two samples.

\noindent
(3) Under further higher rotation speeds multiple MH texture is expected to appear, that is,
CUV. The estimated $\Omega\sim 16$rad/sec for $R=50\mu$m\cite{ishigurothesis} is too high to attain in the
present rotation cryostat in ISSP, whose maximum speed is 11.5rad/s.

\noindent
(4) Our identifications of RD-hb for $R=50\mu$m and RD-ax for $R=115\mu$m are
perfectly matched with the experimental facts that in the former (latter) there does not
exist (do exist) the hysteresis under the $\pm\bf \Omega$ rotations, namely,
the former texture  RD-hb has no polarity and can change continuously under the reversal
of the rotation sense. In contrast, RD-ax has a definite polarity because the $\vec{d}$-vectors point
to the negative $z$ direction for the counter clock-wise rotation, which  never continuously change into
the positive $z$ direction when the rotation sense is reversed, leading to a hysteresis phenomenon.

Because of these facts, we conclude that the un-identified texture at rest and 
lower rotations confined in narrow cylinders is RD texture. Physically
 RD texture with the polar core at the center confined in a narrow cylinder under fields becomes energetically
advantageous over MH, which is characterized by having the A-phase everywhere. 
In order to confirm our identification, we point out several experiments:

\noindent
 (A) The most important prediction is the critical rotation $\Omega_c(T)$
 shown in Fig. \ref{T-Omega} for $R=50\mu$m. As for $R=115\mu$m, $\Omega_c=1.3$rad/s
 at $T/T_c=0.95$ shown in Fig. \ref{Omega-F}. This can be detected by the change of NMR spectrum
 because RD and MH exhibit different spectral features\cite{Ruutu}.

\noindent
 (B) It might be quite interesting to control the texture by tuning the magnetic field $H$
 as shown in Fig. \ref{H-F}. By lowering $H$, MH becomes stable at rest.
 This controllability by $H$ should be utilized to identify textures,
 and furthermore be used for realization of exotic and unexplored physics
 associated with multi-component superfluidity, such as Majorana particle in parallel plates\cite{tsutsumi2,kawakami}.

\noindent
(C) The $\vec l$-vectors in RD-hb and MH-hb, which do not point perfectly to the radial direction,
exhibit a distortion or in-plane twisting as seen from Figs. \ref{RD-hb}(a) and \ref{MH-hb}(a).
This ultimately leads to the mass current along the rotation axis; the $z$-direction.
This non-trivial bending current ($\propto \nabla \times {\vec l}$) should be tested in a future experiment.

\subsection{Supplementary discussion}
Here we discuss the similarity and difference between the present superfluid $^3$He-A and
$p$-wave superfluids in atomic gases\cite{tsutsumi}. In order to gain some perspectives of the present study,
it might be useful to discuss those issues.
The similarity is obvious because both are described by the $p$-wave pairing.
Thus the same GL functional forms, which is derived purely by general symmetry arguments
 can be used for both cases. Depending upon the situations, we can restrict the OP space to effectively
 reduce either orbital degrees of freedom or spin degrees of freedom for a Cooper pair. 
 Thus we can regard the $p$-wave superfluids in atomic gases produced by a field sweep Feshbach 
 resonance as a superfluid $^3$He-A without the spin degrees of freedom because magnetic field polarizes the 
 atom spin, thus the Fermion atoms are spinless.

 One of the main difference between them lies in the boundary condition. 
 $^3$He is usually confined by a rigid wall of a bucket while the atomic gases are trapped by
 a harmonic potential. The former strongly constraints the $\vec l$-vector always perpendicular to a wall
 because the perpendicular motion of a Cooper pair to the wall is suppressed. In the harmonic trap the boundary
 condition acts more gradually and gently. The $\vec l$-vectors tend  to align tangentially at the boundary 
 because the two point nodes along the $\vec l$-vector direction are effectively excluded 
 from the condensate volume region in this configuration, saving the condensation energy.
 Thus the resulting $\vec l$-vector textures in two systems greatly differ. Under rotation the vortices emerged
 are also distinctive\cite{tsutsumi}.

\section*{Acknowledgments}
We acknowledge useful and informative discussions with K. Izumina, R. Ishiguro, M. Kubota, O. Ishikawa, and Y. Sasaki
for experimental aspect of this project and with T. Takagi, T. Ohmi, T. Mizushima, M. Ichioka, and T. Kawakami
for theoretical aspect.
Y. T. is benefitted though the joint research supported by
 the Institute for Solid State Physics, the University of Tokyo.

\end{document}